\documentclass[aps,prd,reprint,superscriptaddress,amsmath,amssymb,nofootinbib,floatfix]{revtex4-2}
\usepackage{graphicx}
\usepackage{dcolumn}
\usepackage[utf8]{inputenc}
\usepackage{multirow}

\usepackage{ulem}
\usepackage{xcolor}
\usepackage[unicode=true,colorlinks,linkcolor=blue,anchorcolor=blue,urlcolor=blue,citecolor=blue,breaklinks=true]{hyperref}
\usepackage{orcidlink}
\usepackage{subfigure}

\newcommand{\ii}{\mathrm{i}\,}

\newcommand{\pararrow}{\mathord{\buildrel{\lower3pt\hbox{$\scriptscriptstyle\leftrightarrow$}}\over {\partial}}} 
\newcommand{\pararrowk}[1]{\mathord{\buildrel{\lower3pt\hbox{$\scriptscriptstyle\leftrightarrow$}}\over {\partial}\hspace*{-0.18em}{}^#1}\hspace*{-0.18em} \,} 
\newcommand{\mytrace}[1]{\langle #1 \rangle} 

\usepackage{physics}
\usepackage{bm}
\usepackage{easyReview}

\newcommand{\YD}{\Upsilon_1(3\,{}^3D_1)}

\newcommand{\qfnu}{\affiliation{College of Physics and Engineering, Qufu Normal University, Qufu 273165, China}}
\newcommand{\hnnu}{\affiliation{Institute of Particle and Nuclear Physics, Henan Normal University, Xinxiang 453007, China}}

\begin{document}
 
	\title{ Hunting for $B\bar B$ molecular state $X_{b0}$ via radiative transition of $\Upsilon(10753)$ }
    
    \author{Yuan-Jun Gao\,} \qfnu
	\author{Gang Li\,\orcidlink{0000-0002-5227-8296}} \email{gli@qfnu.edu.cn} \qfnu
	\author{Shi-Dong Liu\,\orcidlink{0000-0001-9404-5418}} \email{liusd@qfnu.edu.cn}\qfnu
    \author{Qi Wu\,\orcidlink{0000-0002-5979-8569}}\email{wuqi@htu.edu.cn} \hnnu
 
\begin{abstract}
We investigate the radiative decay $\Upsilon(10753) \to \gamma X_{b0}$ within the framework of nonrelativistic effective field theory (NREFT). The $\Upsilon(10753)$ is treated as an $S$-$D$ mixed state of the $\Upsilon(4S)$ and $\YD$, while the $X_{b0}$ is interpreted as a weakly bound $B\bar{B}$ molecule with $J^{PC}=0^{++}$. The decay process was assumed to occur via the intermediate meson loops involving the $S$-wave $B^{(*)}$ and $P$-wave $B_1^{(\prime)}$ mesons. 
Our calculated results indicate that the decay $\Upsilon(10753) \to \gamma X_{b0}$ is dominated by the $B_1^{(\prime)}$ menson loops. 
The partial decay width is predicted to be $0.2-1.5~\mathrm{keV}$ for a binding energy range of $\epsilon_X=0-10~\mathrm{MeV}$, corresponding to a branching fraction of $10^{-6}-10^{-5}$. The decay width is found to be insensitive to the full width of the $B_1^{\prime}$ meson. Our study suggests that the radiative decay of the $\Upsilon(10753)$ is a promising channel to search for the $X_{b0}$ state, which is crucial for testing heavy-quark symmetries and understanding the exotic hadron spectrum in the bottom sector.	
\end{abstract}

\date{\today}

\maketitle

\section{Introduction} \label{sec:intro}
Over the past decades, numerous new hadrons that deviate from the properties expected in the conventional quark model (i.e., $q\bar{q}$ mesons and $qqq$ baryons) have been observed in quarkonium spectra~\cite{ParticleDataGroup:2024cfk}. These hadrons are good candidates for the exotic states~\cite{Godfrey:2008nc,Liu:2013waa,Lebed:2016hpi,Guo:2017jvc,Brambilla:2019esw,Chen:2022asf}, usually referred to collectively as $XYZ$ states. A landmark breakthrough in the study of exotic states occurred in 2003, when the Belle Collaboration discovered the first charmonium exotic state $X(3872)$ in $B \to K \pi^+ \pi^- J/\psi$~\cite{Belle:2003nnu}. Due to the narrow width ($1.19 \pm 0.21~\mathrm{MeV}$), the mass ($3871.64 \pm 0.06~\mathrm{MeV}$) at the $D^0\bar{D}^{*0}$ threshold ($3871.69~\mathrm{MeV}$)~\cite{ParticleDataGroup:2024cfk}, and the quantum numbers $I^G(J^{PC})=0^+(1^{++})$~\cite{LHCb:2013kgk}, the $X(3872)$ is widely interpreted as a loosely bound $D\bar{D}^*$ molecular state~\cite{Godfrey:2008nc,Liu:2013waa,Lebed:2016hpi,Guo:2017jvc,Brambilla:2019esw}. This interpretation has laid the foundation for exploring mesonic molecular systems in the heavy-quark sector. 

In terms of the heavy quark flavor symmetry (HQFS), it is natural to conjecture the existence of the bottomonium counterpart of $X(3872)$~\cite{Hou:2006it,Ozpineci:2013zas,Yang:2017prf,Ding:2020dio,Zhao:2021cvg}, denoted as $X_b$. The $X_b$ has been extensively investigated both theoretically and experimentally (see Refs.~\cite{CMS:2013ygz,ATLAS:2014mka,Guo:2014sca,Li:2014uia,Li:2015uwa,Belle:2014sys,Belle-II:2022xdi,Jia:2023pud,Wang:2023vkx,Liu:2024ets,Belle-II:2025ubm}). Experimentally, the CMS and ATLAS Collaborations searched for the $X_b$ via $X_b\to\Upsilon(1S)\pi^+\pi^-$ in $pp$ collisions~\cite{CMS:2013ygz,ATLAS:2014mka}. Although no clear signal was observed, their analysis suggested that isospin violation in the bottomonium system could suppress the signal strength in certain decay channels. The isospin conserved channel $X_b\to\omega\Upsilon(1S)$ was predicted to have a partial width on the order of tens of keV~\cite{Li:2015uwa}. Moreover, the decay $X_b\to\pi\pi\chi_{bJ}$ is also expected to be experimentally accessible~\cite{Jia:2023pud}. Both tetraquark and molecular state models predict the mass of $X_b$ to lie in the range $10.5\text{--}10.7~\mathrm{GeV}$~\cite{Tornqvist:1993ng,Ebert:2005nc,Matheus:2006xi,Ali:2009pi,Guo:2013sya}. Since the $X_b$ is very heavy and has $J^{PC} = 1^{++}$, a direct discovery at current electron-positron collision facilities is less likely, though Super KEKB may provide an opportunity via $\Upsilon(5S,6S)$ radiative decays~\cite{Aushev:2010bq}. Inspired by the production of $X(3872)$ from highly excited charmonia~\cite{BESIII:2013fnz}, radiative decays of high-lying bottomonia are recognized as one of the most promising production mechanisms. Consequently, decays such as $\Upsilon(5S,6S)\to\gamma X_b$ and $\Upsilon(10753)\to\gamma X_b$ have been theoretically proposed~\cite{Wang:2023vkx,Liu:2024ets}.

Heavy quark spin symmetry (HQSS) further predicts the existence of a complete set of spin partner. In particular, the scalar partner of the $X_b$, a $J^{PC}=0^{++}$ state (called $X_{b0}$ hereafter) is a possible $B\bar{B}$ molecular state near the $B\bar{B}$ threshold. Several theoretical approaches, such as Bethe-Salpeter formalism~\cite{Ke:2012gm,Ding:2020dio}, coupled channel unitary analyses~\cite{Ozpineci:2013zas}, and chiral quark model~\cite{Liu:2008mi,Li:2012mqa}, support the feasibility of such a weakly bound $B\bar{B}$ molecule. However, in contrast to the well studied $X_b$, the production and decay properties of the $X_{b0}$ remain largely unexplored. To our knowledge, the only dedicated theoretical study so far is the work by Britto et al.~\cite{Britto:2024nrp}, which calculated in the molecular picture the $X_{b0}$ production in the radiative decay of the $\Upsilon(4S)$, giving a branching ratio of order $10^{-5}\text{--}10^{-3}$. To better understand the $X_{b0}$, it is helpful to look for other possible production or decay channels.

Recently, the radiative production of exotic molecular states via highly excited heavy quarkonia, especially those with $J^{PC}=1^{++}$, has attracted considerable theoretical attention~\cite{Guo:2013zbw,Wang:2023vkx,Liu:2024ets}. In 2019, the Belle collaboration reported a new structure in a reanalysis of $e^+e^-\to\Upsilon(nS)\pi^+\pi^-(n=1,2,3)$~\cite{Belle:2019cbt}, which is listed in the PDG as $\Upsilon(10753)$ with the quantum numbers $J^{PC}=1^{--}$ and a mass around $10753~\mathrm{MeV}$. This mass provides enough phase space for the radiative production of $X_{b0}$ near the $B\bar{B}$ threshold. Since $\Upsilon(10753)$ can be interpreted as an $S$-$D$ mixture of $\Upsilon(4S)$ and $\YD$~\cite{Li:2021jjt,Li:2022leg,Bai:2022cfz,Belle-II:2023twj}, it has been considered as a promising source for producing $X_b$ in theoretical prediction~\cite{Liu:2024ets}. Alternative interpretations of $\Upsilon(10753)$ also exist, such as conventional bottomonium state~\cite{Chen:2019uzm}, tetraquark~\cite{Wang:2019veq,Ali:2019okl}, or hybrid state~\cite{TarrusCastella:2019lyq}.

In this work, we consider the $X_{b0}$ as a $B\bar{B}$ molecular state and adopt the interpretation of $\Upsilon(10753)$ as a $4S$-$3D$ mixed state. Using nonrelativistic effective field theory (NREFT), we calculate the production of $X_{b0}$ in the radiative decay $\Upsilon(10753)\to\gamma X_{b0}$. Specifically, we include the intermediate meson loop contributions involving both $S$-wave $B^{(*)}$ mesons and $P$-wave $B_1^{(\prime)}$ mesons. Through numerical evaluation, we predict the partial width of this process, analyze the relative importance of different loop contributions, and systematically examine the impact of the finite widths of $B_1^{(\prime)}$.

The rest of this paper is organized as follows. In Sec.~\ref{sec:formula}, we present the theoretical framework. Section~\ref{sec:results} provides the numerical results and discussions. A brief summary is given in Sec.~\ref{sec:summary}.

\section{Theoretical Consideration} \label{sec:formula}

   \subsection{The $4S-3D$ mixing of $\Upsilon(10753)$}
   When classifying the $\Upsilon(10753)$ into the conventional bottomonium family, its experimentally measured mass~\cite{ParticleDataGroup:2024cfk} deviates significantly from the predicted masses of pure bottomonium states~\cite{Badalian:2008ik,Badalian:2009bu,Godfrey:2015dia,Segovia:2016xqb,Wang:2018rjg,Li:2019qsg,Ni:2025gvx}. To address this discrepancy, we adopt the physical interpretation proposed in Refs.~\cite{Li:2021jjt,Bai:2022cfz,Li:2022leg,Luo:2025kid}, in which the $\Upsilon(10753)$ is described as an admixture of the $\Upsilon(4S)$ and $\YD$ states (the partner is the $\Upsilon(10580)$). Within this framework, the $\Upsilon(10753)$ is expressed as the following linear combination:
\begin{align}
\begin{pmatrix}
\Upsilon(10580) \\
\Upsilon(10753)
\end{pmatrix}
=
\begin{pmatrix}
\cos\theta & -\sin\theta \\
\sin\theta & \cos\theta
\end{pmatrix}
\begin{pmatrix}
\Upsilon(4S) \\
\YD
\end{pmatrix}\,,
\end{align}
where $\theta$ is the mixing angle that determines the component proportion of the $4S$ and $3D$ states in the physical states $\Upsilon(10753)$ and $\Upsilon(10580)$. $\Upsilon(4S)$ and $\YD$ describe the wave functions of the pure $4S$ and $3D$ bottomonium states, respectively. 

The value of the mixing angle $\theta$ can be constrained by the experimental data. In Ref.~\cite{Li:2021jjt}, a combined analysis of the measured dielectron width of $\Upsilon(10580)$ and the mixing-induced mass shifts of the $4S$ and $3D$ states determined $\theta$ to be $(33 \pm 4)^\circ$. This indicates that the $\Upsilon(10753)$ is dominated by the $\YD$ component. Following Refs.~\cite{Li:2021jjt,Li:2022leg,Bai:2022cfz}, we also adopt the same value. In this work, we employ the decay widths of $\Upsilon(4S)$ and $\YD$ to bottom meson pairs as calculated from the unquenched quark models in Ref.~\cite{Wang:2018rjg}, which properly account for coupled-channel effects such as S-D mixing and resonance pole shifts. Although the $^3P_0$ model can in principle introduce nonphysical contributions from hard vertices at high momentum, the specific decays considered here occur near threshold and therefore involve only low momentum transfer. Moreover, we adopt only the masses and partial widths from Ref.~\cite{Wang:2018rjg}, with the couplings extracted from leading-order heavy-quark-symmetric Lagrangians, ensuring that no double counting or additional model dependence is introduced.

   \subsection{Intermediate bottom meson loops for $\Upsilon(10753) \to \gamma X_{b0}$}
   
\begin{figure}
	\centering
	\includegraphics[width=0.9\linewidth]{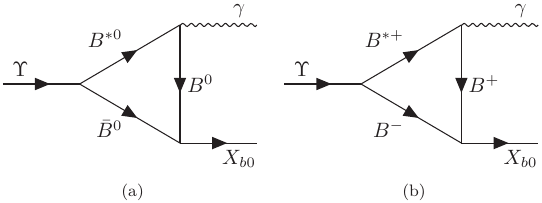}
	\caption{Feynman diagrams for $\Upsilon(10753) \to \gamma X_{b0}$ via $B^{(*)} \bar{B}^{(*)}$ meson loops. The symbol $\Upsilon$ denotes the $\Upsilon(4S)$ and $\YD$ bottomonium states.}
	\label{fig:feynmandiagram1}
\end{figure}
   
Since the $\Upsilon(10753)$ lies above the open-bottom threshold, its dominant decay mode is the strong decay into bottom-antibottom meson pair. Meanwhile, the $X_{b0}$ with $J^{PC}=0^{++}$ is considered as a $B \bar{B}$ molecular state\remove{.}~\footnote{Under the Heavy Quark  Spin Symmtry (HQSS), the $B^*\bar{B}^*$ can also couple into a $0^{++}$ molecular state. However, a $0^{++}$ state near the $B^*\bar{B}^*$ threshold has its pole located on a remote Riemann Sheet~\cite{Baru:2017gwo}, which is far from the physical region or lies on a deeply unphysical sheet. Even if the pole indeed exists theoretically, its coupling to physical open channels (such as $B^*\bar{B}^*$) may be very weak. In this work, we only consider the $B \bar B$($0^{++}$) molecular state, and hence the contribution from the $B^*\bar{B}^*$ molecular state is not included.}. Therefore, the radiative decay of $\Upsilon(10753)$ to $X_{b0}$ can proceed through a triangle meson loop mechanism. In this mechanism, the initial state $\Upsilon(10753)$ first decays into a bottom meson pair and then rescatter into the final state $X_{b0}$ and photon by exchanging a bottom meson. In Fig.~\ref{fig:feynmandiagram1}, we show the contributions from the $B^{(*)} \bar{B}^{(*)}+\mathrm{c.c.}$ intermediate meson loops. Since the coupling of $\Upsilon(10753)$ to $B_s\bar{B}_s$ is expected to be greatly weak~\cite{Liang:2019geg}, the contributions from the loops involving $B_s$ mesons are not considered. 

\begin{figure}
	\centering
	\includegraphics[width=0.9\linewidth]{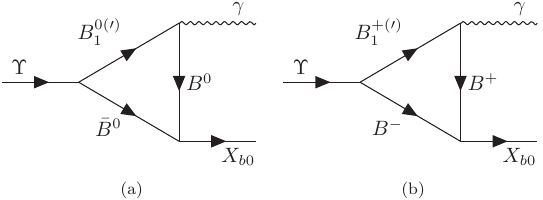}
	\caption{Feynman diagrams for $\Upsilon(10753) \to \gamma X_{b0}$ via the $B_1^{(\prime)}\bar{B}^{(*)}$ meson loops.}
	\label{fig:feynmandiagram2}
\end{figure}

The $X_{b0}$ with $J^{PC}=0^{++}$ couples to a $B\bar{B}$ meson pair in an $S$-wave. Apart from the $P$-wave couplings to the two $S$-wave bottom mesons as shown in Fig. \ref{fig:feynmandiagram1}, the initial $1^{--}$ bottomonium and the final-state photon can also couple to one $S$-wave and one $P$-wave bottom meson in an $S$-wave as shown in Fig.~\ref{fig:feynmandiagram2}. The power counting analysis in the following section indicates that the loops in Fig.~\ref{fig:feynmandiagram2} provide the dominant contribution to the radiative decay $\Upsilon(10753) \to \gamma X_{b0}$, which confirmed by our numerical calculations. 

   \subsection{The effective Lagrangians}
In heavy quark effective theory, bottom mesons are classified by the total angular momentum $s_\ell^P$ of their light degrees of freedom. For each value of $s_\ell^P$, the corresponding degenerate doublet can be described uniformly by a superfield~\cite{Casalbuoni:1996pg,Hu:2005gf}. In this work, we employ nonrelativistic effective field theory to calculate the decay widths. We use the symbols $H$, $S$, and $T$ to describe the bottom mesons with $s_\ell^P={1/2}^-$, ${1/2}^+$, and $3/2^+$, respectively~\cite{Casalbuoni:1996pg,Hu:2005gf},
\begin{align}
    H_a &= \vec{V}_a\cdot\vec{\sigma} + P_a\,,\\
    S_a &= \vec{V}_{1a}^\prime\cdot\vec{\sigma} + P_{0a}\,,\\
	T^i_a &= V_{2a}^{ij}\sigma^j+\sqrt{\frac{2}{3}} V_{1a}^i + \ii \frac{1}{\sqrt{6}}\epsilon^{ijk} V_{1a}^j\sigma^k\,.
\end{align}
Here, $\vec{\sigma}$ is the Pauli matrix. $V_a=(B^{*0},\,B^{*+})$ and $P_a=(B^0,\,B^+)$ denote the vector and pseudoscalar bottom mesons with $s_\ell^P={1/2}^-$, respectively. The $s_\ell^P={1/2}^+$ doublet is represented by $V_{1a}^\prime=(B_1^{\prime0},\,B_1^{\prime+})$ and $P_{0a}=(B_0^0,\,B_0^+)$. For $s_\ell^P={3/2}^+$, we have $V_{1a}=(B_1^0,\,B_1^+)$ and $V_{2a}=(B_2^0,\,B_2^+)$. The $V_{2a}$ does not contribute in the present calculations and is therefore omitted. The corresponding fields for the charge-conjugated mesons are given by~\cite{Casalbuoni:1996pg,Hu:2005gf}
\begin{align}
   \bar{H}_a &= -\vec{\bar{V}}_a\cdot\vec{\sigma} + \bar{P}_a\,,\\
    \bar{S}_a &= - \vec{\bar{V}}_{1a}^\prime\cdot\vec{\sigma} + \bar{P}_{0a}\,,\\
	\bar{T}^i_a &= -\bar{V}_{2a}^{ij}\sigma^j+\sqrt{\frac{2}{3}} \bar{V}_{1a}^i - \ii \frac{1}{\sqrt{6}}\epsilon^{ijk} \bar{V}_{1a}^j\sigma^k\,. 
\end{align}

In the heavy quark limit, the leading order Lagrangians describing the couplings of $S$-wave and $D$-wave bottomonia to bottom-antibottom meson pairs are given by~\cite{Casalbuoni:1996pg,Hu:2005gf}
\begin{align}
	\mathcal{L}_{\Upsilon(4S)} &= \ii \frac{g_S}{2} \mytrace{\bar{H}_a^\dagger \vec{\sigma} \cdot \pararrow H_a^\dagger J}\nonumber\\
        &+ \frac{g_S^\prime}{2} \mytrace{(\bar{H}_a^\dagger S_a^\dagger + \bar{S}_a^\dagger H_a^\dagger) J} + \mathrm{H.c.}\,,\label{eq:LagS}\\
        \mathcal{L}_{\Upsilon(3D)} &= \ii \frac{g_D}{2} \mytrace{\bar{H}_a^\dagger \sigma^i \pararrowk{j} H_a^\dagger J^{ij}}\nonumber\\
	&+ \frac{g_D^\prime}{2} \mytrace{(\bar{T}_a^{\dagger j} \sigma^i H_a^\dagger - \bar{H}_a^{\dagger}\sigma^i T_a^{\dagger j}) J^{ij}} +  \mathrm{H.c.}\,,\label{eq:LagD}
\end{align}
where $A \pararrow B = A (\partial B) - (\partial A) B$. The $S$-wave bottomonium field is represented by the combination $J = \vec{\Upsilon} \cdot \vec{\sigma} + \eta_b$, while the $D$-wave bottomonium $\YD$ is described by the tensor field $J^{ij} = \frac{\sqrt{15}}{10} (\Upsilon_1^i \sigma^j + \Upsilon_1^j \sigma^i) - \frac{1}{\sqrt{15}} \delta^{ij} \vec{\Upsilon}_1 \cdot \vec{\sigma}$~\cite{Guo:2013zbw}. $g_S [g_D]$ and $g_S^\prime [g_D^\prime]$ are the coupling constants of $\Upsilon(4S) [\YD]$ to a ${1/2}^-$ -- ${1/2}^-$ pair and a ${1/2}^-$ -- ${1/2}^+ [{3/2}^+]$ pair of bottom mesons, respectively. $a$ is the light flavor index. The symbol $\langle \cdots \rangle$ denotes a trace in spinor space, and a dagger ($\dagger$) indicates the outgoing particle in the relevant vertex.

Utilizing the effective Lagrangians above and the theoretical decay widths for $\Upsilon(4S) \to B\bar{B}$ and $\YD \to B\bar{B}$, $B\bar{B}^* + \mathrm{c.c.}$, $B^*\bar{B}^*$ from Ref.~\cite{Wang:2018rjg}, we can extract the corresponding coupling constants. For the $S$-wave, we obtain $g_S = 0.776~\mathrm{GeV^{-3/2}}$. For the D-wave, the coupling $g_D$ takes values $0.157~\mathrm{GeV^{-3/2}}$, $0.376~\mathrm{GeV^{-3/2}}$, and $1.879~\mathrm{GeV^{-3/2}}$. Furthermore, applying the ratios $g_S^\prime/g_S = 1.5~\mathrm{GeV}$ and $g_D^\prime/g_D = 1.7~\mathrm{GeV}$ derived from Ref.~\cite{Liu:2024ets}, we determine the corresponding couplings $g_S^\prime$ and $g_D^\prime$.

We now consider the vertices involving the $X_{b0}$. In Ref.~\cite{Gao:2025zhp}, the charmonium-like $X_0$ is interpreted as a $D\bar{D}$ bound state. Based on heavy quark symmetry, $X_{b0}$ can be regarded as a $B\bar{B}$ molecular state. Assuming equal contributions from the charged and neutral components, the wave function of this state can be written as
\begin{equation}
	\ket*{X_{b0}} = \frac{1}{\sqrt{2}} (\ket*{B^0 \bar{B}^0} + \ket*{B^+ B^-})\,.
\end{equation}

The coupling of the $X_{b0}$ to the bottom-antibottom meson pairs is given by
\begin{align}\label{eq:LagXb0}
	\mathcal{L}_X = \frac{1}{\sqrt{2}} X_{b0}^\dagger (g_1 B^0 \bar{B}^0 + g_2 B^+ B^-) + \mathrm{H.c.}\, ,
\end{align}

The coupling constants $g_i$ ($i=1,2$) are related to the binding-energy relations established in Refs.~\cite{Weinberg:1965zz,Baru:2003qq}:
\begin{equation}\label{eq:couplegi}
	g_i = \left(\frac{16\pi}{\mu_B} \sqrt{\frac{2\epsilon_X}{\mu_B}}\right )^{1/2}\,,
\end{equation}
where $\mu_B = m_B m_{\bar{B}}/(m_B + m_{\bar{B}})$ is the reduced mass and $\epsilon_X = m_B + m_{\bar{B}} - M_{X_{b0}}$ is the binding energy. It should be emphasized that Eq.~\eqref{eq:couplegi} is merely a model-dependent estimate and only valid for a shallow bound state to guarantee a well-controlled EFT expansion. In the deeply bound regime, higher-order terms and short-range dynamics, which are not considered therein, may introduce significant theoretical uncertainties. According to the criterion for a shallow bound state discussed in Ref.~\cite{Guo:2017jvc}, the binding energy should satisfy $E_b<1/(2\mu R^2_{\rm{conf}})$, where $E_b$ is the binding energy, $\mu$ is the reduced mass, and $R_{\rm{conf}}$ is the confinement radius. Hence, according to this condition, when the $X_{b0}$ binding energy is smaller than $7.4~\mathrm{MeV}$, it could be safely regarded as a loosely bound state. Hence, the range of $\epsilon_X$ in the present work is taken as $\sim 0-10~\mathrm{MeV}$.

The effective Lagrangian describing the radiative transition of $S$-wave bottom mesons with photon emission reads~\cite{Amundson:1992yp,Hu:2005gf,Guo:2013zbw,Liu:2024ets}
\begin{equation}\label{eq:LagphotonS}
	\mathcal{L}_{HH\gamma} = \frac{e\beta}{2} \mytrace{H_a^\dagger H_b \vec{\sigma} \cdot \vec{B} Q_{ab}} + \frac{eQ^\prime}{2m_{Q^\prime}} \mytrace{H_a^\dagger \vec{\sigma} \cdot \vec{B} H_a}\,,
\end{equation}
where $B^k = \epsilon^{ijk} \partial^i A^j$ is the magnetic field. The first term contains the light quark charge matrix $Q = \mathrm{diag}(-1/3,\,2/3)$, while the second term involves the heavy quark electric charge $Q^\prime$ and mass $m_{Q^\prime}$.

The effective Lagrangian $\mathcal{L}_{S/TH\gamma}$ describing the interactions of the $s_\ell^P={1/2}^+$ or ${3/2}^+$ bottom mesons, photons, and $s_\ell^P={1/2}^-$ bottom mesons is given by~\cite{Guo:2013zbw,Wang:2023vkx,Liu:2024ets}:
\begin{equation}\label{eq:LagphotonP}
	\mathcal{L}_{S/TH\gamma} = -\frac{\ii e \tilde{\beta}}{2} \mytrace{H_a^\dagger S_b \vec{\sigma} \cdot \vec{E} Q_{ab}} +  \mytrace{T_a^i H_b^\dagger C_{ab}} E^i\,.
\end{equation}
Here $E^i$ is the electric field and  $C$ is a  $2\times 2$ diagonal matrix that parameterizes the coupling strength in the second term.

The magnetic coupling parameters $\beta$, $\tilde{\beta}$, and $C_{ab}$ for bottom mesons are determined from the radiative decay processes described by the expanded Lagrangians in Eqs.~\eqref{eq:LagphotonSexpand} and \eqref{eq:LagphotonPexpand}. The relations between the partial widths and the coupling constants are given by
\begin{align}
   \Gamma(B^* \to \gamma B) &= \frac{1}{12 \pi M_{B^*}}|\frac{e Q^\prime}{m_{Q^\prime}} + e\beta Q_{ab}|^2 E_\gamma^3 M_B\,,\\
   \Gamma(B_1^\prime \to \gamma B) &= \frac{1}{8 \pi M_{B_1^\prime}}|e \tilde{\beta} Q_{ab}|^2 E_\gamma^3 M_B\,,\\
   \Gamma(B_1 \to \gamma B) &=\frac{1}{3 \pi M_{B_1}}|C_{ab}|^2 E_\gamma^3 M_B \,,
\end{align}
where the masses $M_{B_1^\prime}=5739.20~\mathrm{MeV}$ and $M_{B_1}=5755.12~\mathrm{MeV}$ are taken from the potential model predictions in Ref.~\cite{Asghar:2018tha}. Using these masses together with the partial widths calculated in Refs.~\cite{Zhu:1996qy,Choi:2007se,Asghar:2018tha}, we obtain $\beta = 2.12~\mathrm{GeV^{-1}}$, $\tilde{\beta} = 1.84~\mathrm{GeV^{-1}}$, and $C=\mathrm{diag}(-0.12~\mathrm{GeV^{-1}},\,0.22~\mathrm{GeV^{-1}})$.

The explicit expanded forms of the above effective Lagrangians are provided in Appendix~\ref{appendixA}. Using these Lagrangians, we have derived the decay amplitudes for the processes shown in Figs.~\ref{fig:feynmandiagram1} and \ref{fig:feynmandiagram2}. The complete expressions are given in Appendix~\ref{appendixB}. The partial decay width is
\begin{equation}
	\Gamma(\Upsilon(10753)\to\gamma X_{b0}) = \frac{E_\gamma |\mathcal{M}_{\Upsilon(10753)\to\gamma X_{b0}}|^2}{24 \pi M_{\Upsilon(10753)}^2} \, ,
\end{equation}
where $E_\gamma$ is the photon energy in the $\Upsilon(10753)$ rest frame.

The $1/2^+$ bottom meson $B_1^\prime$ is predicted to have a large width. In Ref.~\cite{Du:2017zvv} a large width of about $238~\mathrm{MeV}$ was predicted, whereas the prediction in Ref.~\cite{Asghar:2018tha} show a small value of $126~\mathrm{MeV}$. The latter work also provides a width prediction of $16.4~\mathrm{MeV}$ for the $3/2^+$ $B_1$ meson, agreeing with the
measured data between $27.5\text{--}31~\mathrm{MeV}$~\cite{ParticleDataGroup:2024cfk}. This large width effect for $B_1^{(\prime)}$ was taken into account
in our calculations by using the Breit–Wigner parameterization
to approximate the spectral function of the $B_1^{(\prime)}$. The corresponding mass-distribution function is defined as
\begin{equation}
    \rho_{B_1^{(\prime)}}(s) = \frac{1}{\pi} \frac{M_{B_1^{(\prime)}}\Gamma_{B_1^{(\prime)}}}{(s-M_{B_1^{(\prime)}}^2)^2+M_{B_1^{(\prime)}}^2\Gamma_{B_1^{(\prime)}}^2}\,,
\end{equation}
where $s$ is the mass squared of ${B_1^{(\prime)}}$ and $M_{B_1^{(\prime)}}$ is the central mass of $B_1^{(\prime)}$. Then the
amplitude is given by~\cite{Wang:2023vkx,Wu:2018xaa,Liu:2024ets}
\begin{equation}
\mathcal{M}_{B_1^{(\prime)}} =\frac{1}{W_{B_1^{(\prime)}}}\int_{s_l}^{s_h} \rho_{B_1^{(\prime)}}(s) \bar{\mathcal{M}}_{B_1^{(\prime)}}(M_{B_1^{(\prime)}} \to \sqrt{s})ds\,,
\end{equation}
where $W_{B_1^{(\prime)}} = \int_{s_l}^{s_h}\rho_{B_1^{(\prime)}}(s)ds$ is the normalization factor with \textcolor{blue}{$s_l=M_B^2$} and $s_h=(M_{B_1^{(\prime)}}+\Gamma_{B_1^{(\prime)}})^2$. $\bar{\mathcal{M}}_{B_1^{(\prime)}}(M_{B_1^{(\prime)}} \to \sqrt{s})$ represents the loop amplitude of $B_1^{(\prime)}$ calculated using
$s$ as the mass squared. The lower and upper limits of the integral are based on a reasonable truncation determined by the decay threshold and the spectral function of the broad $B^\prime_1$ resonance, which is approximated by the Breit-Wigner parametrization~\cite{Wu:2018xaa, Wang:2023vkx}.

\section{Numerical results}
\label{sec:results}

To assess the relative importance of the different bottom meson loop contributions, we perform a power counting analysis in the nonrelativistic framework, using the velocity $v \ll 1$ of the intermediate mesons as the expansion parameter. This method has been applied to study intermediate mesons loop effects~\cite{Guo:2010ak,Guo:2010zk,Guo:2013zbw,Wu:2018xaa,Wang:2023vkx,Liu:2024ets}. In the power counting scheme, the momentum scales as $v$, and the kinetic energy scales as $v^2$. 

For the diagrams in Fig.~\ref{fig:feynmandiagram1}, the initial bottomonium couples to the $S$-wave bottom mesons in a $P$-wave. This momentum contracts with the external photon momentum $q$ from the photon vertex and thus should be counted as $q$. The vertex involving $X_{b0}$ is an S-wave coupling and introduces no additional momentum or velocity factor. The vertices involving the photon are also in a $P$-wave, which should be counted as $q$. Consequently, the decay amplitude for Fig.~\ref{fig:feynmandiagram1} scales as
\begin{equation}\label{eq:counting1}
N \frac{v^5}{(v^2)^3} \frac{q^2}{m_B^2} = N \frac{E_\gamma^2}{m_B^2 v}\,,
\end{equation}
where $v$ is the average of intermediate bottom meson velocities $v_1$ and $v_2$ at the two different cuts in the loop diagram~\cite{Guo:2012tg}, estimated as $v_1=\sqrt{|m_1 + m_2 - M_i|/\bar{m}_{12}}$ and $v_2=\sqrt{|m_2 + m_3 - M_f|/\bar{m}_{23}}$. The value $v \approx 0.08 \text{--} 0.11$. The factor $N$ denotes the product of the coupling constants from the three interaction vertices. To render the amplitude dimensionally consistent, the factor $q^2$ is accompanied by $m_B^{-2}$, where $m_B$ is the bottom meson mass.

For the loops in Fig.~\ref{fig:feynmandiagram2}, all the vertices in the $B_1^{(\prime)}\bar{B}^{(*)}$ meson loops  are momentum-independent $S$-wave couplings. Therefore, the corresponding amplitude scales as
\begin{equation}\label{eq:counting2}
N^\prime \frac{v^{\prime\,5}}{(v^{\prime\,2})^3}\frac{E_\gamma}{m_B} = N^\prime \frac{E_\gamma}{m_B v^\prime}\,.
\end{equation}
Here, the energy $E_\gamma$ originates from the $S$-wave photon vertex in the Lagrangians [Eqs.~\eqref{eq:LagphotonSexpand} and \eqref{eq:LagphotonPexpand}]. The quantity $v^\prime$ is the average velocity of the $B_1^{(\prime)}\bar{B}^{(*)}$ meson loops, which falls in the range $v^\prime \simeq 0.11 \text{--} 0.13$.

Since the coupling constants in  Figs.~\ref{fig:feynmandiagram1} and \ref{fig:feynmandiagram2} are of the same order (i.e., $N \sim N^\prime$), the relative contribution of Figs.~\ref{fig:feynmandiagram1} and \ref{fig:feynmandiagram2} is estimated as:
\begin{equation}
r = \frac{m_B v}{v^\prime E_\gamma} \approx 21.0\,.
\end{equation}

The above power counting result demonstrates that the $B_1^{(\prime)}\bar{B}^{(*)}$ meson loops in Fig.~\ref{fig:feynmandiagram2} provide a significantly larger contribution to the radiative decay than those in Fig.~\ref{fig:feynmandiagram1}. We note that the values of $v$, $v^\prime$, and the ratio $r$ obtained above span a range because the mass of $X_{b0}$ has been varied within $M_{X_{b0}} = 10.549\text{--}10.559~\mathrm{GeV}$, corresponding to a binding energy $\epsilon_X = 0 \text{--} 10~\mathrm{MeV}$.

\begin{figure}
	\centering
	\includegraphics[width=0.98\linewidth]{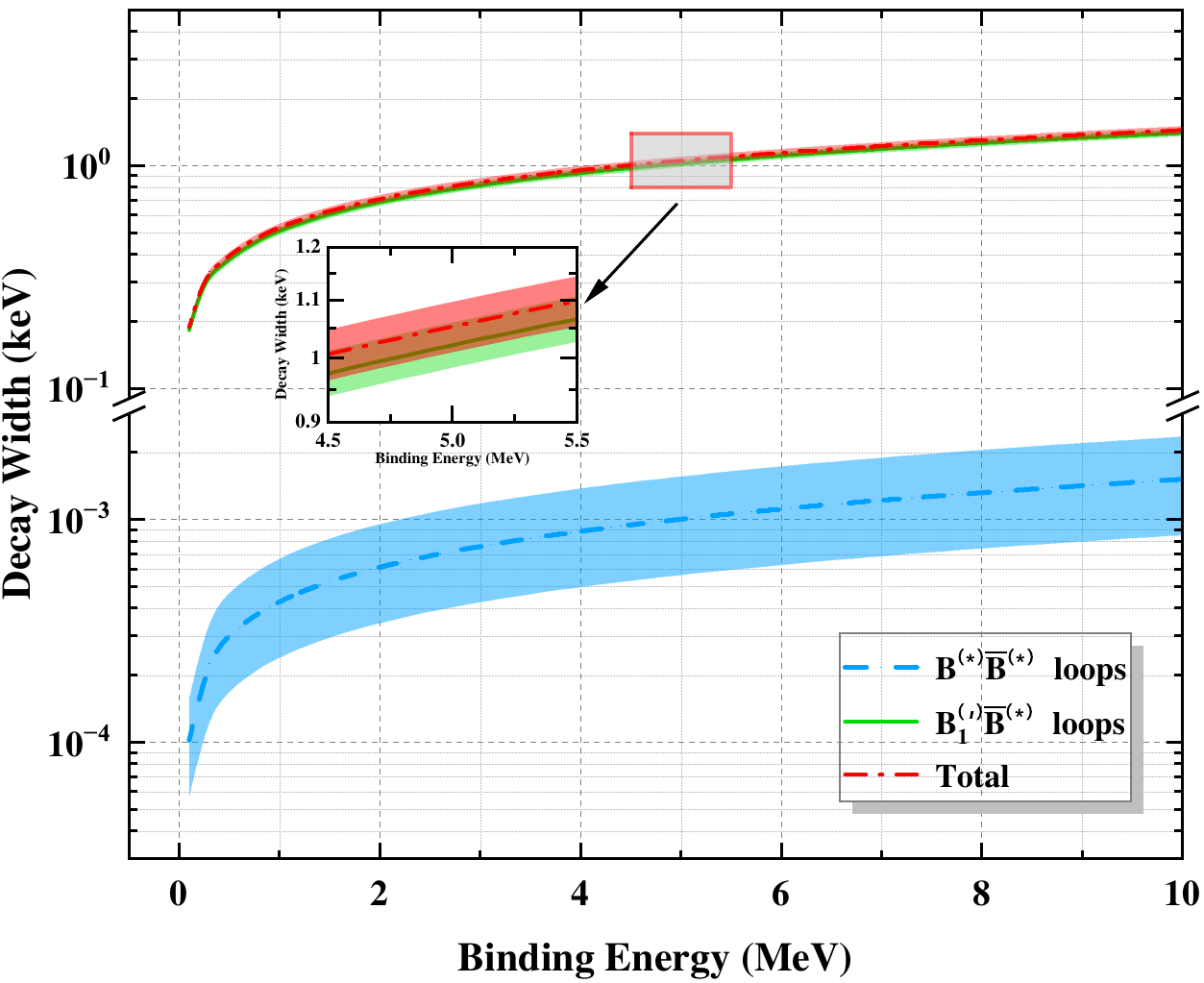}
	\caption{Partial widths of the radiative decay $\Upsilon(10753) \to \gamma X_{b0}$ contributing from different meson loops as a function of the binding energy $\epsilon_X$ with $\Gamma_{B_1^{(\prime)}} = 0$. The light-colored bands represent the uncertainties arising from the $4S$-$3D$ mixing angle $\theta$.}
	\label{fig:Y10753toXb0}
\end{figure}

In Fig.~\ref{fig:Y10753toXb0}, we present the partial width of $\Upsilon(10753) \to \gamma X_{b0}$  as a function of the binding energy from $0$ to $10~\mathrm{MeV}$. The partial width increases monotonically with $\epsilon_X$. Across the entire considered range, the contribution from the $B_1^{(\prime)}\bar{B}^{(*)}$ meson loops in Fig.~\ref{fig:feynmandiagram2} is significantly larger than that from the $B^{(*)}\bar{B}^{(*)}$ meson loops in Fig.~\ref{fig:feynmandiagram1}, which is consistent with our previous expectations based on power counting analysis. For a certain binding energy, the errors arising from the $4S$-$3D$ mixing angle $\theta$ is about $\mathcal{O}(0.05)$. The partial decay width is approximately proportional to the the coupling constant square. 


\begin{figure}
	\centering
	\includegraphics[width=0.98\linewidth]{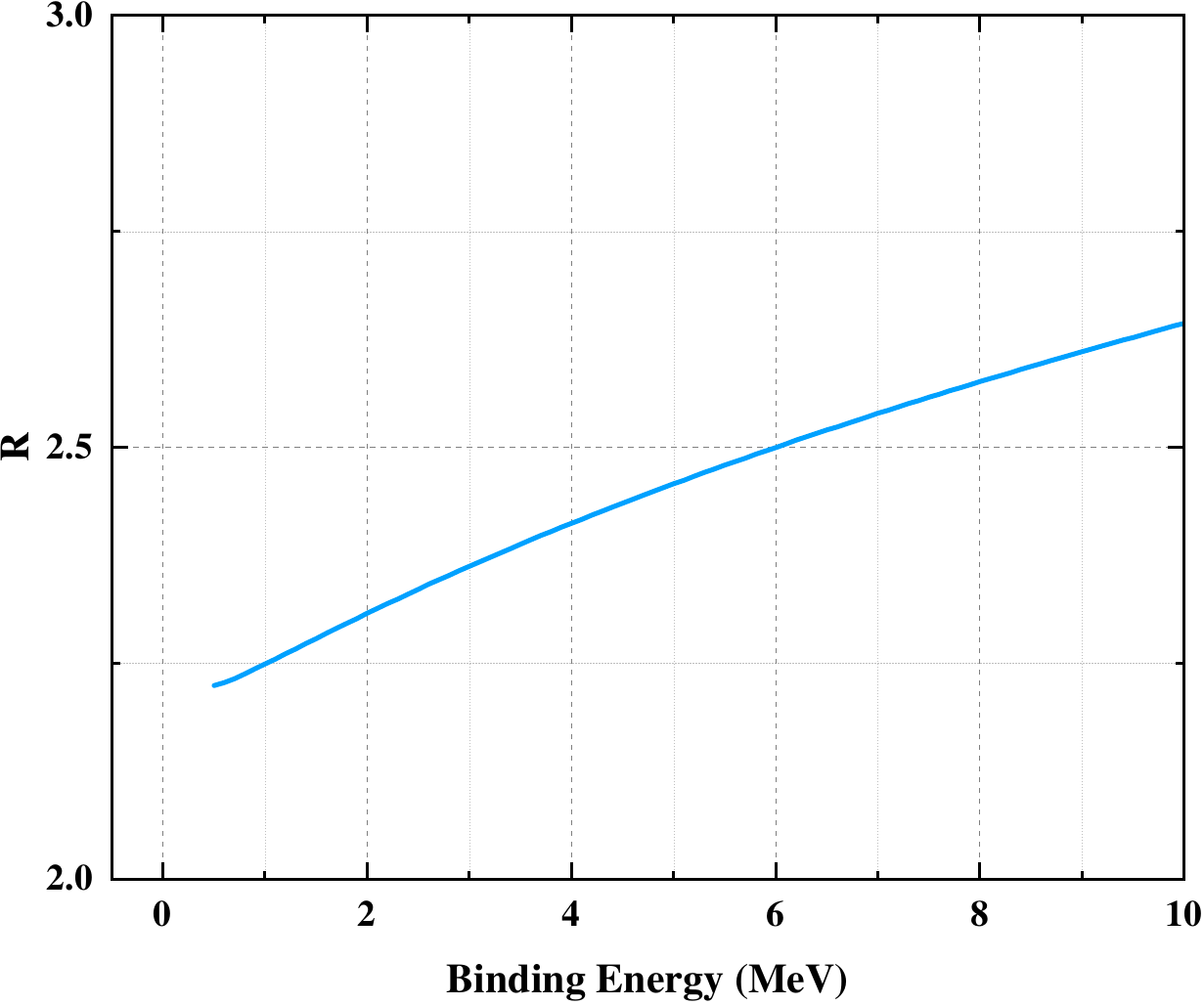}
	\caption{Dependence of the ratio $R=\mathcal{B}[\Upsilon(10753) \to \gamma X_b]/\mathcal{B}[\Upsilon(10753) \to \gamma X_{b0}]$ on the binding energy $\epsilon_X$.}
	\label{fig:ratio}
\end{figure}

Furthermore, incorporating the results of Ref.~\cite{Liu:2024ets}, we define the ratio $R=\mathcal{B}[\Upsilon(10753) \to \gamma X_b]/\mathcal{B}[\Upsilon(10753) \to \gamma X_{b0}]$. In Fig.~\ref{fig:ratio}, we plot the dependence of this ratio on the binding energy. It is seen
that although the binding energy dependence exists, it is actually
weakened strongly in comparison to the absolute decay width shown in Fig.~\ref{fig:Y10753toXb0}.

\begin{table*}[htbp]
	\caption{The predicted decay partial width (in units of keV) of $\Upsilon(10753) \to \gamma X_{b0}$ for different binding energies. Here we choose $\Gamma_{B_1^\prime}=0, 100, 200, 300~\mathrm{MeV}$.}
	\label{tab:effectwidth}
	\begin{ruledtabular}
		\begin{tabular}{lccccccccc}
  \multirow{2}{*}{$\epsilon_X$} & \multirow{2}{*}{$B^{(*)}\bar{B}^{(*)}$ loops} & \multicolumn{4}{l}{$B_1^{(\prime)}\bar{B}^{(*)}$ loops} & \multicolumn{4}{l}{Total decay width}\\
  \cline{3-6} \cline{7-10}
  ~ & ~ & $\Gamma_{B_1^\prime}=0$ & $\Gamma_{B_1^\prime}=100$ & $\Gamma_{B_1^\prime}=200$ & $\Gamma_{B_1^\prime}=300$ & $\Gamma_{B_1^\prime}=0$ & $\Gamma_{B_1^\prime}=100$ & $\Gamma_{B_1^\prime}=200$ & $\Gamma_{B_1^\prime}=300$\\ 
  \colrule
  $1~\mathrm{MeV}$ & $4.278\substack{+2.373\\-1.883} \times 10^{-4}$ & $0.509\substack{+0.019\\-0.018}$ & $0.534\substack{+0.024\\-0.022}$ & $0.512\substack{+0.025\\-0.022}$ & $0.485\substack{+0.025\\-0.021}$ & $0.527\substack{+0.023\\-0.022}$ & $0.552\substack{+0.027\\-0.026}$ & $0.528\substack{+0.028\\-0.026}$ & $0.500\substack{+0.027\\-0.024}$\\
  $2~\mathrm{MeV}$ & $6.116\substack{+3.392\\-2.692} \times 10^{-4}$ & $0.685\substack{+0.026\\-0.025}$ & $0.720\substack{+0.033\\-0.030}$ & $0.693\substack{+0.034\\-0.030}$ & $0.659\substack{+0.034\\-0.029}$ & $0.708\substack{+0.030\\-0.030}$ & $0.742\substack{+0.036\\-0.034}$ & $0.713\substack{+0.037\\-0.034}$ & $0.678\substack{+0.036\\-0.032}$\\
  $5~\mathrm{MeV}$ & $1.005\substack{+0.557\\-0.442} \times 10^{-3}$ & $1.022\substack{+0.039\\-0.037}$ & $1.075\substack{+0.049\\-0.044}$ & $1.040\substack{+0.052\\-0.046}$ & $0.996\substack{+0.051\\-0.044}$ & $1.053\substack{+0.044\\-0.043}$ & $1.105\substack{+0.052\\-0.050}$ & $1.068\substack{+0.055\\-0.051}$ & $1.021\substack{+0.053\\-0.048}$\\
  $10~\mathrm{MeV}$ & $1.518\substack{+0.842\\-0.668} \times 10^{-3}$ 
  & $1.404\substack{+0.054\\-0.051}$ & $1.478\substack{+0.067\\-0.061}$ & $1.438\substack{+0.071\\-0.064}$ & $1.383\substack{+0.071\\-0.062}$ & $1.442\substack{+0.059\\-0.058}$ & $1.514\substack{+0.070\\-0.067}$ & $1.470\substack{+0.073\\-0.068}$ & $1.414\substack{+0.072\\-0.066}$\\
		\end{tabular}
	\end{ruledtabular}
\end{table*}

In Table~\ref{tab:effectwidth}, we list the contributions of $\Upsilon(10753) \to \gamma X_{b0}$ from $B^{(*)}\bar{B}^{(*)}$ loops, $B_1^{(\prime)}\bar{B}^{(*)}$ loops, and the total contributions at several illustrative values $\epsilon_X=1, 2, 5, 10~\mathrm{MeV}$.
For the $B_1^\prime$, we choose the $\Gamma_{B_1^\prime}$ to be 0, 100, 200, and 300 MeV, respectively. It can be seen that the contributions from $B^{(*)}\bar{B}^{(*)}$ loops are about $10^{-3}$ keV. For $B_1^{(\prime)}\bar{B}^{(*)}$ loops, the decay width first increases and then decreases with the increase of the $B_1^\prime$ width, which can reach several keVs. 

\begin{figure}
	\centering
	\includegraphics[width=0.98\linewidth]{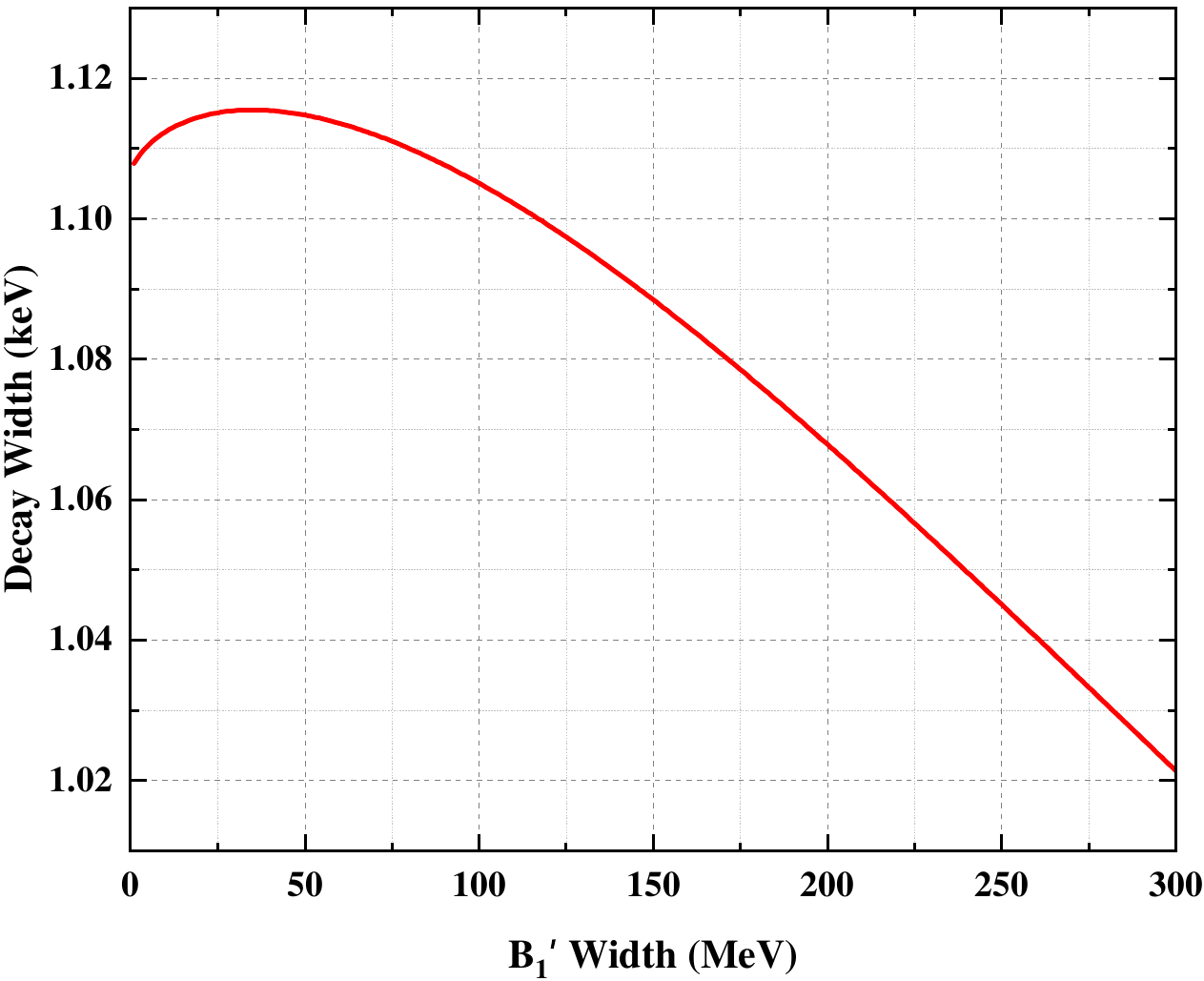}
	\caption{Dependence of the partial width as a function of the ${B_1^\prime}$ width. The $B_1$ width is fixed at $25~\mathrm{MeV}$ and the binding energy is $\epsilon_X = 5~\mathrm{MeV}$.}
	\label{fig:B1primewidth}
\end{figure}

In Fig.~\ref{fig:B1primewidth}, we show the $B_1^\prime$ width dependence of the radiative decay width for the $\Upsilon(10753) \to \gamma X_{b0}$. In our calculations, the $B_1$ width is fixed to be $25~\mathrm{MeV}$ due to its smallness in comparison with that of the $B_1^\prime$. It is seen that the radiative decay width exhibits only a weak dependence on the width of the $B_1^\prime$. Within the considered range, its variation falls into $1.02–1.12~\mathrm{keV}$. The decay width first increases and then decreases with the increase of the $B_1^\prime$ width and reach their maximum around $\Gamma_{B_1^{\prime}}\simeq 35~\mathrm{MeV}$. It should be noted that the possible variation of the couplings resulting from the change of the $B_1^{(\prime)}$ width is not considered in our calculations, which might give rise to extra effect on the radiative decay width.

\section{Summary}
\label{sec:summary}
In this work, we investigate the radiative decays $\Upsilon(10753)\to\gamma X_{b0}$ using nonrelativistic effective field theory. In the calculations, the $X_{b0}$ is assumed to be a molecular state of $B\bar{B}+\mathrm{c.c.}$ with $J^{PC}=0^{++}$ and the $\Upsilon(10753)$ is an $S-D$ mixed state of the $\Upsilon(4S)$ and $\YD$. The numerical calculations were performed under two kinds of intermediate bottomed meson loops. The first kind is $B^{(*)}\bar{B}^{(*)}$ loop coupled with $\Upsilon(10753)$ in $P$-wave and the second is $B_1^{(\prime)}\bar{B}^{(*)}$ loop coupled with $\Upsilon(10753)$ in S-wave.

It is found that the possible effect of the large width of the $B_1^\prime$ meson on the radiative decay width might be of minor importance, if the couplings do not change substantially with the $B_1^{(\prime)}$ width. For a binding energy range of $0-10~\mathrm{MeV}$, the partial decay width of this reaction has been estimated to be of the order of $0.2-1.5~\mathrm{keV}$, corresponding to a branching fraction of $10^{-6}-10^{-5}$. This appreciable of the radiative width implies that searching for the $X_{b0}$ via the process $\Upsilon(10753)\to\gamma X_{b0}$ is promising. The observation of the $X_{b0}$ might help us to study the nature of the $X_b$ partner states related by the heavy quark symmetry.

\begin{acknowledgments}
\label{sec:acknowledgements}
This work is supported by the National Natural Science Foundation of China under Grants No. 12475081, 12235018 and 12405093, as well as supported, in part, by National Key Research and Development Program under Grant No. 2024YFA1610504; by Taishan Scholar Project of Shandong
Province; and by the Natural Science Foundation of Shandong Province under Grant No. ZR2025MS04.
\end{acknowledgments}

\onecolumngrid
\begin{appendix}
\section{Lagrangians expansion}\label{appendixA}

The Lagrangians in Eqs.~\eqref{eq:LagS}, \eqref{eq:LagD}, \eqref{eq:LagphotonS} and \eqref{eq:LagphotonP} are expanded as
    \begin{align}\label{eq:LagSexpand}
	\mathcal{L}_{\Upsilon(4S)} &= \ii g_S \Upsilon^i (\bar{V}_a^{\dagger j} \pararrowk{i} V_a^{\dagger j} - \bar{V}_a^{\dagger j} \pararrowk{j} V_a^{\dagger i} - \bar{V}_a^{\dagger i} \pararrowk{j} V_a^{\dagger j}) - g_S \epsilon^{ijk} \Upsilon^i (\bar{V}_a^{\dagger k} \pararrowk{j} P_a^\dagger + \bar{P}_a^\dagger \pararrowk{j} V_a^{\dagger k}) +\ii g_S \Upsilon^i \bar{P}_a^\dagger \pararrowk{i} P_a^\dagger\nonumber\\
	&+ \epsilon^{ijk} g_S \eta_b \bar{V}_a^{\dagger i} \pararrowk{j} V_a^{\dagger k} + \ii g_S \eta_b (\bar{P}_a^\dagger \pararrowk{i} V_a^{\dagger i} - \bar{V}_a^{\dagger i} \pararrowk{i} P_a^\dagger)
	+ g_S^\prime \eta_b (\bar{P}_{0a}^\dagger P_a^\dagger + \bar{P}_a^\dagger P_{0a}^\dagger ) - g_S^\prime \eta_b (\bar{V}_{1a}^{\prime \dagger i} V_a^{\dagger i} + \bar{V}_a^{\dagger i} V_{1a}^{\prime\dagger i})\nonumber\\
	&+ g_S^\prime \Upsilon^i (\bar{P}_a^\dagger V_{1a}^{\prime\dagger i} + \bar{P}_{0a}^\dagger V_a^{\dagger i} - \bar{V}_a^{\dagger i} P_{0a}^\dagger - \bar{V}_{1a}^{\prime\dagger i} P_a^\dagger)
	+ \ii g_S^\prime \epsilon^{ijk} \Upsilon^i (\bar{V}_a^{\dagger k} V_{1a}^{\prime\dagger j} + \bar{V}_{1a}^{\prime\dagger k} V_a^{\dagger j})+\mathrm{H.c.}\,,
    \end{align}

    \begin{align}\label{eq:LagDexpand}
	\mathcal{L}_{\Upsilon(3D)} &= \ii g_D \frac{\sqrt{15}}{3} \Upsilon_1^i \bar{P}_a^\dagger \pararrowk{i} P_a^\dagger
	+ g_D \frac{\sqrt{15}}{6} \epsilon^{ijk} \Upsilon_1^i (\bar{P}_a^\dagger \pararrowk{j} V_a^{\dagger k} + \bar{V}_a^{\dagger k} \pararrowk{j} P_a^\dagger)+ \ii \frac{ g_D}{2\sqrt{15}} \Upsilon_1^i (4 \bar{V}_a^{\dagger j} \pararrowk{i} V_a^{\dagger j} - \bar{V}_a^{\dagger j} \pararrowk{j} V_a^{\dagger i} - \bar{V}_a^{\dagger i} \pararrowk{j} V_a^{\dagger j})\nonumber\\
	&+ g_D^\prime \frac{\sqrt{10}}{2} \Upsilon_1^i (\bar{V}_{1a}^{\dagger i} P_a^\dagger - \bar{P}_a^\dagger V_{1a}^{\dagger i})+ \ii g_D^\prime \frac{\sqrt{10}}{4} \epsilon^{ijk} \Upsilon_1^i (\bar{V}_{1a}^{\dagger k} V_a^{\dagger j} + \bar{V}_a^{\dagger k} V_{1a}^{\dagger j}) + \mathrm{H.c.}\,,
    \end{align}

    \begin{align}\label{eq:LagphotonSexpand}
	\mathcal{L}_{HH\gamma} &= \ii e \beta Q_{ab} \partial^i A^j (V_a^{\dagger i} V_b^j - V_a^{\dagger j} V_b^i)
	+ e \beta Q_{ab} \epsilon^{ijk} \partial^i A^j (P_a^\dagger V_b^k + V_a^{\dagger k} P_b)\nonumber\\
	&+ \ii \frac{e Q^\prime}{m_{Q'}} \partial^i A^j (V_a^{\dagger j} V_a^i - V_a^{\dagger i} V_a^j) + \frac{e Q^\prime}{m_{Q'}} \epsilon^{ijk} \partial^i A^j (V_a^{\dagger k} P_a + P_a^\dagger V_a^k)\,,
    \end{align}

    \begin{align}\label{eq:LagphotonPexpand}
	\mathcal{L}_{S/TH\gamma} &= e \tilde{\beta} Q_{ab} \epsilon^{ijk} \partial^0 A^i V_a^{\dagger j} V_{1b}^{\prime k}- \ii e \tilde{\beta} Q_{ab} \partial^0 A^i (V_a^{\dagger i} P_{0b} + P_a^\dagger V_{1b}^{\prime i})\nonumber\\
	&+ \ii \sqrt{\frac{2}{3}} C_{ab} \epsilon^{ijk} \partial^0 A^i V_{1a}^j V_b^{\dagger k}+ 2 \sqrt{\frac{2}{3}} C_{ab} \partial^0 A^i P_b^\dagger V_{1a}^i\,.
    \end{align}

    \section{Transition Amplitudes}\label{appendixB}
    In the rest frame of initial particle, the transition amplitude for $\Upsilon(4S)(p) \to \left[B^*(l)\bar{B}(p-l)\right]B(l-q) \to \gamma(q)X_{b0}(p-q)$ reads
    \begin{align}
    \mathcal{M}_{4S} &= \frac{\sqrt{2}}{2} g_i g_S(e \beta Q_{ab} + \frac{e Q^\prime}{m_{Q'}})[2q \cdot \varepsilon(-q) q \cdot \varepsilon(p)- 2q \cdot q\varepsilon(p) \cdot \varepsilon(-q)] I^{(1)}(M_{B^*}\,,M_{B}\,,M_{B}\,,\vec{q}\,)\,.
    \end{align}
The amplitude for $\YD(p) \to \left[B^*(l)\bar{B}(p-l)\right]B(l-q) \to \gamma(q)X_{b0}(p-q)$ reads
\begin{align}
    \mathcal{M}_{3D} &= -\frac{1}{2} \sqrt{\frac{5}{6}} g_i g_D(e \beta Q_{ab} + \frac{e Q^\prime}{m_{Q'}})[2q \cdot \varepsilon(-q) q \cdot \varepsilon(p)
    - 2q \cdot q\varepsilon(p) \cdot \varepsilon(-q)] I^{(1)}(M_{B^*}\,,M_{B}\,,M_{B}\,,\vec{q}\,)\,.
\end{align}

Since $\Upsilon(10753)$ is a mixed state of $\Upsilon(4S)$ and $\YD$, the amplitude for $\Upsilon(10753)(p) \to \left[B^*(l)\bar{B}(p-l)\right]B(l-q) \to \gamma(q)X_{b0}(p-q)$ in Fig.~\ref{fig:feynmandiagram1} is given by
\begin{equation}
	\mathcal{M}_{Fig.1} = 2\sum\left(\mathcal{M}_{4S} \sin{\theta} + \mathcal{M}_{3D} \cos{\theta} \right )\,,
\end{equation}
where the factor 2 accounts for charge conjugation, and the summation runs over amplitudes from both neutral and charged bottom meson intermediate states.

The amplitude for $\Upsilon(4S)(p) \to \left[B_1^\prime(l)\bar{B}(p-l)\right]B(l-q) \to \gamma(q)X_{b0}(p-q)$ reads
\begin{equation}
    \mathcal{M}_{4S}^\prime = \frac{\sqrt{2}}{2} g_i g_S^\prime e \tilde\beta Q_{ab} E_\gamma \varepsilon(p) \cdot \varepsilon(-q) I(M_{B_1^\prime}\,,M_{B}\,,M_{B}\,,\vec{q}\,)\,.
\end{equation}
The amplitude for $\YD(p) \to \left[B_1(l)\bar{B}(p-l)\right]B(l-q) \to \gamma(q)X_{b0}(p-q)$ reads
\begin{equation}
    \mathcal{M}_{3D}^\prime = - \ii \sqrt{\frac{10}{3}} g_i g_D^\prime C_{ab} E_\gamma \varepsilon(p) \cdot \varepsilon(-q) I(M_{B_1}\,,M_{B}\,,M_{B}\,,\vec{q}\,)\,. 
\end{equation}

The amplitude corresponding to the process $\Upsilon(10753)(p) \to \left[B_1^{(\prime)}(l)\bar{B}(p-l)\right]B(l-q) \to \gamma(q)X_{b0}(p-q)$ shown in Fig.~\ref{fig:feynmandiagram2}  is written as 
\begin{equation}
	\mathcal{M}_{Fig.2} = 2\sum\left(\mathcal{M}_{4S}^\prime \sin{\theta} + \mathcal{M}_{3D}^\prime \cos{\theta} \right )\,.
\end{equation}

It should be noted that these amplitudes must be multiplied by the appropriate non-relativistic normalization factor $\sqrt{M_{\Upsilon(10753)} M_{X_{b0}}}m_1 m_2 m_3$. $m_i$ ($i=1,2,3$) represents the mass of the intermediate meson.

The scalar three-point loop integral associated with the triangle diagram in Fig.~\ref{fig:feynmandiagram2} follows from Ref.~\cite{Guo:2010ak} and is given by
\begin{align}
    I(m_1\,,m_2\,,m_3\,,\vec{q}\,)
    &= \ii \int \frac{d^d l}{(2\pi)^d} \frac{1}{(l^2-m_1^2+\ii \epsilon)[(p-l)^2-m_2^2+\ii \epsilon][(l-q)^2-m_3^2+\ii \epsilon]}\nonumber\\
    & \approx \frac{\mu_{12} \mu_{23}}{16\pi m_1 m_2 m_3}\frac{1}{\sqrt{a}}\{\arctan(\frac{c^\prime-c}{2\sqrt{ac}}) + \arctan[\frac{2a+c-c^\prime}{2\sqrt{a(c^\prime-a)}}]\}\,,
\end{align}
where $\mu_{ij}$ are the reduced masses, $b_{12} = m_1 + m_2 - M, b_{23} = m_2 + m_3 + q^0 - M$ with $M$ being the mass of the initial particle, and 
\begin{equation}
a = {(\frac{\mu_{23}}{m_3})}^2 \vec{q}^{\,2}, \quad  c = 2\mu_{12}b_{12},  \quad  c^\prime = 2\mu_{23}b_{23} + \frac{\mu_{23}}{m_3} \vec{q}^{\,2}\,.
\end{equation}

The vector loop integral $I^{(1)}(m_1\,,m_2\,,m_3\,,\vec{q}\,)$ in the decay amplitudes for Fig.~\ref{fig:feynmandiagram1} is~\cite{Guo:2010ak}
\begin{align}
    q^i I^{(1)}(m_1\,,m_2\,,m_3\,,\vec{q}\,)
    = \ii \int \frac{d^d l}{(2\pi)^d} \frac{l^i}{(l^2-m_1^2+\ii \epsilon)[(p-l)^2-m_2^2+\ii \epsilon][(l-q)^2-m_3^2+\ii \epsilon]}\,,
\end{align}

\begin{align}
    I^{(1)}(m_1\,,m_2\,,m_3\,,\vec{q}\,) \approx \frac{\mu_{23}}{a m_3}[B(c^\prime-a)-B(c)
    + \frac{1}{2}(c^\prime-c)I(m_1\,,m_2\,,m_3\,,\vec{q}\,)]\,.
\end{align}
Here the function $B(c)$ is
\begin{equation}
B(c) = - \frac{\mu_{12}\mu_{23}}{4m_1 m_2 m_3}\frac{\sqrt{c-i\epsilon}}{4\pi}\,.
\end{equation}

\end{appendix}

\twocolumngrid   
\bibliography{references.bib}
\end{document}